\documentclass[acmsmall]{acmart}

\makeatletter
\def\runningfoot{\def\@runningfoot{}}
\def\firstfoot{\def\@firstfoot{}}
\makeatother
\renewcommand\footnotetextcopyrightpermission[1]{} % removes footnote with conference information in first column
\usepackage{etoolbox}
\makeatletter
\def\@copyrightspace{\relax}
\makeatother
\makeatletter
\let\@authorsaddresses\@empty
\makeatother
\usepackage{graphicx}
\usepackage{capt-of}
\usepackage{multirow}

\acmVolume{}
\acmNumber{}
\acmArticle{}
\usepackage{colortbl}
\usepackage{pdflscape}
\begin{document}

\title{Organizational Bulk Email Systems: Their Role and Performance in Remote Work}

\author{Ruoyan Kong}
\email{kong0135@umn.edu}
\affiliation{%
  \institution{University of Minnesota - Twin Cities}
%   \city{Twin Cities}
  \state{USA}
}

\author{Haiyi Zhu}
\email{haiyiz@cs.cmu.edu}
\affiliation{%
  \institution{Carnegie Mellon University}
%   \city{City}
  \state{USA}
}

\author{Joseph A. Konstan}
\email{konstan@umn.edu}
\affiliation{%
  \institution{University of Minnesota - Twin Cities}
%   \city{City}
  \state{USA}
}


\begin{abstract}
The COVID-19 pandemic has forced many employees to work from home. Organizational bulk emails now play a critical role to reach employees with central information in this work-from-home environment.
However, we know from our own recent work that organizational bulk email has problems: recipients fail to retain the bulk messages they received from the organization; recipients and senders have different opinions on which bulk messages were important; and communicators lack technology support to better target and design messages.
In this position paper, first we review the prior work on evaluating, designing, and prototyping organizational communication systems. Second we review our recent findings and some research techniques we found useful in studying organizational communication. Last we propose a research agenda to study organizational communications in remote work environment and suggest some key questions and potential study directions.

\end{abstract}

% <concept_id>10003120.10003130.10011762</concept_id>
% <concept_desc>Human-centered computing~Empirical studies in collaborative and social computing</concept_desc>

% \ccsdesc[500]{Human-centered computing~Empirical studies in collaborative and social computing}

% \keywords{organizational communication, email, remote work}

\maketitle

\pagestyle{plain}
\section{Introduction}

Hundreds of millions of people around the world have been working from home due to the COVID-19 pandemic. As organizations are switching to remote-work, organizational communication is becoming more important than ever. Organizations need to  announce pandemic-related news, remote-work policies, work arrangements, and health information to their employees. 
During the COVID-19 pandemic, these messages have to meet  rapidly changing circumstances and conditions. 
Also, the organizational communicates largely rely on telecommunication technologies such as emails, given that many conventional in-person communication channels are no longer available.
%(entities comprising multiple people with a particular purpose \cite{handy2007understanding}) 

%Remote work refers to organizational work that is performed outside of the normal organizational confines of space and time \cite{10.1145/358061.358068}. 

We know from our own work and prior research that organizational communication system has problems in traditional non-remote-work environment: messages are overwhelming \cite{waller2012impact}; recipients are receiving emails irrelevant to them while missing important ones \cite{doi:10.1080/02680513.2018.1556090} \cite{Jackson:2003:UEI:859670.859673}; recipients are not reading the messages, they just ``open and close'' the messagese; senders and recipients have different perspectives on message values \cite{economic_model,intelligent}; organizations are fragmented in the responsibilities of communication; communicators lack technology to support them with the message design and distribution, and are sending messages out at the cost of communication channels' credibility and effectiveness \cite{kong2023towards}. 

% Specifically, in the remote-work situation, there might be additional problems. Negotiation becomes more difficult \cite{10.1145/302979.303067}.Managers need to communicate extensively with their subordinates to proceed organizational tasks \cite{10.1145/358061.358068}. Leadership performance will be influenced by the remote communication effectiveness \cite{neufeld2010remote}.

It is even more challenging to create effective and efficient organizational communication systems as organization switches to remote work. For example, negotiation about work plans, task assignments, and term meanings between co-workers, managers and employees might become more difficult. Online communication channels are not sufficient when managers have to communicate complicated situations with their subordinates to accomplish organizational tasks. Text-based messages might be difficult to convey the sense of connectivity, which becomes more important in the remote-work environment. In this paper, we try to identify what are the questions needed to be answered in studying organizational communication system in remote work environment, both from our own recent work and prior research.

In this position paper, we first review our recent work and prior research on evaluating, designing, and prototyping organizational communication systems. We then present research techniques we found useful in studying organizational bulk communication. Finally, we propose a research agenda on studying remote work organizational communications with key questions and potential study directions.

\section{Review of Prior Work}
\subsection{Organizational Communication}
Communication within organizations has been studied for more than a century \cite{van2012electronic}. Communication has been called \textit{``the life blood of organization''} \cite{goldhaber1990organizational}, \textit{``the glue that binds it all together''} \cite{katz2008communication}, \textit{``the organization embalming fluid''} \cite{kreps1990game}.  Myers and Myers \cite{myers1982managing} defined \textbf{organizational communications} as “the central binding force that permits coordination among people and thus allows for organized behavior''. Stohl and Redding defined organizational communication as the collective and interactive process of generating and interpreting messages within organizations to achieve their purposes  \cite{stohl1995organizational}.

%\subsection{Mechanisms of Organizational Communication}
Studies on organizational communication's mechanisms have been influenced by sociology, psychology, rhetoric, anthropology, and even the physical sciences \cite{miller2008organizational}. The base of communication mechanism was often traced to the information theory proposed by Claude Shannon (1949) \cite{shannon1948mathematical}, in which he split the communication process in to information source, transmitter, and receiver.
% see Fig \ref{fig:comm}.
% \begin{figure}[!htbp]
% \centering
%   \includegraphics[width=0.8\columnwidth]{figures/shannon.png}
%   \caption{Communication as a mechanistic system. }~\label{fig:comm}
% \end{figure}
Baker in his book \textit{``Organizational Communication''} \cite{baker2007organizational} categorized organizational communication into vertical communication (between hierarchically positioned persons), horizontal communication (between persons who do not stand in hierarchical relation to one another), and diagonal communication (between managers and workers located in different functional divisions). Greenbaum \cite{greenbaum1974audit} defined \textbf{purpose of organizational communications} as the achievement of organizational goals, accomplished through the appropriate employment of communication networks, communication policies, and communication activities.

%\subsection{Evaluation of Organizational Communication}

Scholars proposed different models for evaluating organizational communication's effectiveness. Robert and O'Reilly \cite{roberts1974measuring} proposed using an organizational communication questionnaire (OCQ), which gathered information on trust, influence, mobility, desire for interaction from employees. Stohl and Redding \cite{stohl1987messages} considered employees just an opportunity for cognitive failure in a communication process. Greenbaum, however, argued that communication effectiveness had to be measured by looking at the overall communication system with the activities of employees \cite{greenbaum1974audit}.

\subsection{Remote-Work Organizational Communication}
\subsubsection{Remote-Work Organizational Communication with Productivity}
The probability of remote work was discussed as early as 1980s with the development of telecommunication technology \cite{10.1016/0378-7206(84)90041-7}. A key issue in remote work and virtual organizational structures is the communication between employees who are located remotely and their manager. Olson \cite{10.1145/358061.358068} surveyed a company experimenting with pilot work-at-home
programs in that time. Olson found that in the remote-work situation, the availability of communications was seen as critical; remote workers
needed to be easy to reach within a reasonable amount of time. Barness et al. \cite{motivation} surveyed an Internet commerce firm and found that remote work would reduce the sense of connectedness (which they called as social network centrality), thereby decreased job-focused impression management. Staples \cite{Staples} surveyed 1,343 remote-working individuals in 18 North American organizations and found that too much communication might decrease remote-working productivity; managers of remote employees should focus on activities that demonstrate competence, responsibility and professionalism. Kraut et al. \cite{kraut2002understanding} compared the differences between local communication and remote-work communication, found that remote conversation were more difficult to initiated and extended, misunderstandings were more difficult to be repaired.

\subsubsection{Remote-Work Organizational Communication Technology}
A lot of new technologies have been developed for remote work: e-mail, bulletin boards, instant messaging, document sharing, video conference, awareness services \cite{kraut2002understanding}. Remote-working employees had different levels of satisfaction and productivity with these technologies under different situation. Olson and Meader's survey \cite{olson1997face} with remote-working employees showed that the quality of work with remote high-quality video is as good as face-to-face. Remote work without video was not as good as face-to-face. Conversely, Veinott et al. compared the performance and communication of people explaining a map route to each other \cite{10.1145/302979.303067}, found that video made people more satisfied with the work, but it did not help the quality of the work. Firari \cite{firari2007email} proposed that email communication styles would influence the email interpretation of remote-working employees. Which technology should be used given different messages and remote-working situations is an unknown problem.

\subsection{Email as an Organizational Communication Channel}

\subsubsection{Email for Remote Work}
The widely-application of emails brought changes to organizational communication. Email enabled sender and recipients to control the timing of their portion of the communication, speed up the exchange of information and leaded to the exchange of new information \cite{sproull1986reducing}, expedited communication frequency \cite{feldman1987electronic}, created what Sproull and Kiesler called a “networked organization” \cite{sproull1991computers} in which people can be available when they are physically absent. 

\subsubsection{Email Overload}
The widely-application of emails in workplace also brought email overload. Whittaker and Sidner’s seminal article “Email Overload” \cite{whittaker1996email} used the term to refer not to people being burdened with too much email, but rather to people using email for multiple purposes, i.e., overloading its use. Waller and Ragsdell \cite{waller2012impact} surveyed employees from a multinational service organisation finding particular harm to work-life balance. Merten and Gloor \cite{merten2010too} found that employee job satisfaction went down as internal email volume increased. 

% \subsection{Organizational Email System as a Multi-Stakeholder Problem}
% Bulk email communications within organizations is not a problem only about recipients, but also an example of a multi-stakeholder problem \cite{andriof2002introduction} --- the stakeholders in organizations include:
% \begin{itemize}
%   \item Senders: the original producer of messages, who are the communicators' internal clients.
%   \item Communicators: the staff who are in charge of designing and distributing bulk emails.
%   \item Recipients: the employees who receive bulk emails from the communicators.
% \end{itemize}
%  Not only are there different goals for senders, communicators, and recipients, but also there is a key fourth stakeholder --- the organization itself --- that has its priorities not always recognized by senders, communicators or recipients.  Senders and communicators naturally focus on their own needs --- getting the word out, establishing evidence of notice or compliance, or preserving a record of communication. But recipients, faced with more communication than they can handle, have to scan, filter, or simply ignore messages. In turn, organizational goals around compliance, informed employees, and employee productivity may suffer.
 
 \subsubsection{Bulk Email}
 Bulk email is email that is sent to a large group of recipients \cite{doi:10.1080/02680513.2018.1556090}. Organizations often use bulk emails to deliver messages to their employees \cite{overload}. Example organizational bulk emails include announcements of new staff, summaries of meetings, health and safety issues, and event invitations to relevant groups within organizations, etc. Most work about bulk email were about the emails outside of organizations. Trespalacios and Perkins \cite{trespalacios2016effects} examined the effects of mass email designs (different survey invitation conditions) on response rate, finding that neither the degree of personalization nor the length of the invitation email impacted survey response or completion. Al-Jarrah et al. proposed header-based approaches \cite{al2012identifying}, reached over $90\%$ ROC in CEAS2008 dataset.
 
 In the next section, we presented our recent work about studying a specific channel of organizational communication --- organizational bulk emails. We identified the importance of studying from multi-stakeholders' perspectives in conducting organizational communication research. We also provided the research techniques we found useful in studying organizational communication.

\section{A Case Study of an Organizational Bulk Email System}
%\subsection{Study From Multi-Stakeholders' Perspectives}

We recently conducted a case study on organizational bulk email systems, using a multi-stakeholder perspective \cite{kong2020learning}. Note that the study was conducted at a large organization before it switches to remote work.

The biggest difference between studying communication within and outside organizations is that with organizational context, whether an employee should get a message is not equal to whether this employee likes it, because sometimes he or she has responsibility for knowing about this message though he or she might not have interest in it. And this communication system relies on communication professionals --- communicators' efforts to design and distribute message. 

Therefore, bulk email communications within organizations is not a problem only about recipients, but also an example of a multi-stakeholder problem \cite{andriof2002introduction} --- the stakeholders include:
\begin{itemize}
   \item Senders: the original producer of messages, who are the communicators' internal clients.
   \item Communicators: the staff who are in charge of designing and distributing bulk emails.
   \item Recipients: the employees who receive bulk emails from the communicators.
\end{itemize}
And there is a key fourth stakeholder --- the organization itself --- that has its priorities not always recognized by senders, communicators or recipients.  Senders and communicators naturally focus on their own needs --- getting the word out and establishing evidence of notice or compliance. But recipients, faced with more communication than they can handle, might filter, ignore, scan, or ``open and close'' messages. In turn, organizational goals around compliance, informed employees, and employee productivity may suffer. Therefore, we identified a need to study organizational communication from a whole organization's view with multi-stakeholders' perspectives.

\subsection{Research Methods}
We conducted an in-depth case study of a representative organization --- University of Minnesota, which is a large university, with hierarchical organizational structures, centralized and decentralized communication offices. We reported the research techniques we found useful in studying organizational communication here.

\subsubsection{Engaging stakeholders in the study design}
To get the domain knowledge of the organizational communication system, first we met with a group of 9 communicators within this university to get their opinions on our research questions, potential participants, and knowledge on the structure of the organizational communication system. We found agreement on that the current organizational bulk email system might not be what the university wanted it to be, too often they felt that their messages were ignored. We kept meeting with this communicator group through out research. They supported us to move forward.

In the first meeting, we worked with them to identify the structure of the bulk email system of the university. Fig \ref{fig:comm} in the appendix is the structure of the bulk email system of the university.

% Each working unit had a communication office. There were centralized communication offices and decentralized communication offices. The centralized communication offices sometimes sent bulk emails directly to employees by existing mailing lists, or pulling out lists from Salesforce's database; other times they used a distribution mechanism where the bulk emails were sent to the decentralized communication offices, and asked them to distribute the messages to their employees selectively. Besides sending bulk emails themselves, communicators also send content to newsletter editors. The content would be put into the templates designated by newsletter editors.

% Then we identified three categories of organizational emails: 
% \begin{itemize}
%     \item \textit{Organizational Non-Bulk Emails:} All messages from within the university not sent through a bulk email mechanism. E.g., individual messages from other employees.
%     \item \textit{Organizational Mass Emails:} The organizational bulk emails with a single message. E.g., an email sent to all employees for announcing a free financial workshop.
%     \item \textit{Organizational Newsletter Emails:} A special kind of organizational bulk email that has a collection of messages and is sent to the recipients periodically. E.g., an email sent to all employees daily about the events happening in the university today.
% \end{itemize}

\subsubsection{Quantitative study --- eliminating the influence of message content}
A difficulty of studying organizational communication quantitatively is that the communication is not only influenced by the communication channels, but also the message content itself. We reported here how we eliminated the influence of the message content in a survey on email effectiveness across this university. We worked with the communicators to verify our survey meet the following criteria:
\begin{itemize}
    \item The survey questions are paired: a real message and a corresponding fake message with similar content features (actionability, importance level, relevancy), to test whether the recipients could recognize the real messages from the fake messages.
    \item The message pairs have different channel features: newsletter or single email, position in the email, whether from leadership, etc.
    \item The real messages were received by all participants and have general importance to all employees (thus employees should know about them from the university's perspective).
\end{itemize}

% We asked the participants to indicate whether they could recall the message on a 5 point likert scale where 1 is ``I Have Not Seen it'' and 5 is ``I Have Seen it''. We considered a score of 4 or 5 on a message from a participant as the participant claimed that he or she had seen this message otherwise he or she had not seen it. At the beginning of the survey, we informed that there were both real and fake messages (without a specific number of each type of message). At the end of the survey, we showed the participants which of the messages were real and which were fake.

Then the percentage of participants who claimed they had seen fake messages is the population who did not ``read'' messages but only remembered they usually received similar content. Thus we could define the effectiveness of a real message as $\% \ real \ message \ claimed \ seen - \% \ corresponding \ fake \ message \ claimed \ seen$, without the influence of message content.

% We compared the recognizing ratio between the corresponding real message and fake message. 

% We defined the participant pool as employees who had received the real messages we selected in the past 2 weeks before the survey was distributed, were not senior leaders of the university, and were part of the university's volunteer pool for conducting usability studies on university systems. We randomly selected 3000 participants from the pool and got 162 completed responses.

\subsubsection{Qualitative study --- walking through participants' inboxes}
A good data resource we found in studying organizational communication is the inbox data. We reported how we used inbox data to study the practices and experience with organizational bulk emails of different stakeholders in artifact walkthroughs \cite{10.1145/1013115.1013124} with communicators, recipients, and recipients' managers within the university. 

% We worked with the same group of communicators to stratified sample a list including communicators, recipients with different working units, titles, and length of work experience, excluding senior leaders like unit heads.

% 6 communicators, 9 recipients (5 non-communicator staff and 4 faculty) agreed to participate in our interview. After interviewing the recipients, we asked them whether they would like to participate in an optional study where we invite their direct manager and discuss the non-personal emails we collected from them. Two recipients agreed to join this study and invited their managers.

% \noindent\textbf{A. Interview with Communicators}
% We used inbox data to study the practice of communicators in detail. 

For each communicator, we wanted to reproduce her designing and targeting process of emails from her inbox. We asked her to find important and unimportant emails from her point of views from her inbox. Then we asked how these emails were designed, what are their goals and target population, how were they sent out and measured.  

For each recipient, we wanted to reproduce her reactions with the emails received from both inbox logged data and inbox-review data. Firstly, she was asked to copy and paste 10 email queries to select different subsets of emails she received in the past week. For example, \textit{newer\_than:7d,in:anywhere AND NOT from:me AND label:unread AND from:umn.edu} showed the emails the participant received from the university and unread within 1 week in a Gmail inbox. We used these queries to retrieve the number of emails she received of each type. Second, she was asked to identify the bulk emails she received in the past week from her inbox. Then she recalled how she dealt with these emails and why she chose to deal in that ways.

For each manager, we wanted to simulate her reaction with her employees' email practice with the non-personal emails we collected from the recipients' inbox. The answers from the employee-manager pair are confidential from each other. We showed the manager the bulk emails we collected from her employee, and asked she to give their preferred actions that the employee has done with those emails --- unread or opened. 

In this way, we compared different stakeholders' opinions from the inbox data.

% We could also collect the text data by letting participants forwarding discussed emails to us.

% \noindent\textbf{B. Recipients}

% Interviews with recipients collected both inbox logged data and inbox-review qualitative data. Firstly, they were asked to copy and paste 10 email queries (see Table \ref{tab:command}) to select different subsets of emails they received in the past week. These queries were pretested in one of the author's inbox to ensure they retrieve the right subset of emails. We used these queries to retrieve the number of emails they received of each type. 

% In the second step, the participants were asked to identify the bulk emails and newsletter emails they received in the past week, answer specific questions (see appendix) about how they dealt with these emails and why they chose to deal in that ways. We also asked them to forward these emails to us. These self-reported inbox-review data offered us different results with the inbox log-data when we studied them, as many recipients only ``open and close'' emails.

% \noindent\textbf{C. Managers}

% We used the non-personal emails we collected from the recipients' inbox here. The answers from the employee-manager pair are confidential from each other. We showed them the bulk emails we collected from their employees, and ask them to give their preferred actions that their employees have done with those emails --- unread or opened. In this way we could compare different stakeholders' opinions with the same email.\\

\subsection{Results}
In this section, we report the main takeaways of the study.

\subsubsection{Email overload.} Participants (recipients) reported that they, in general, received too many organizational emails. From recipients' inbox logged data, faculty received 175 organizational emails a week (148.5 organizational non-bulk emails, 26.8 organizational bulk emails and for staff, these numbers are 102.6 and 33.8.
\subsubsection{Recipients were not reading and retaining emails.}
The recipients did not retain most messages in organizational bulk emails, though they opened many of them. From the inbox logged data, $72.92\%$ organizational bulk emails were reported being opened by faculty and only $12.5\%$ of them were read in detail; for staff is $59.18\%$ and $24.49\%$. From our survey, the real messages were claimed seen by only $38\%$ participants, and the fake messages were also claimed seen by $16\%$ participants. That suggested a $22\% = 38\% - 16\%$ average effectiveness.
\subsubsection{Disconnect between senders and recipients over what's important.} Communicators felt they were sending important messages through organizational bulk emails and had good performance while recipients disagreed. Changes in the benefits, leadership, public safety, and administration were recognized as the emails with general importance by communicators and managers. Communicators usually sent them to all faculty and staff. On another side, recipients sometimes felt these emails from university leaders were too high level to be related to themselves thus they did not them. The level of bulk email is defined in Table~\ref{tab:sender_level}. Figure~\ref{fig:open_level} shows the average open rate and read in detail rate of emails from different level's emails. It shows that the higher the email's level, the lower the probability that the email will be read. 
\begin{table}[ht]\vspace*{-0.2in}
\begin{minipage}[b]{0.45\linewidth}
\centering
    % \begin{table}[
    % \centering
    \begin{tabular}{|c|c|c|} 
    \hline
     \textbf{\#}  & \textbf{Level}  & \textbf{Meaning}                                                               \\ 
    \hline
    1             & University      & \begin{tabular}[c]{@{}c@{}}Sent from university-\\wide offices. \end{tabular}  \\ 
    \hline
    2             & College         & \begin{tabular}[c]{@{}c@{}}Sent from college-\\wide offices. \end{tabular}     \\ 
    \hline
    3             & Department      & \begin{tabular}[c]{@{}c@{}}Sent from department-\\wide offices. \end{tabular}  \\
    \hline
    \end{tabular}
    \caption{The definition of email's levels. For example, university-wide offices indicates that the office was in charge of sending bulk emails to recipients across the university.}
    \label{tab:sender_level}
    % \end{table}
\end{minipage}\hfill
\begin{minipage}[b]{0.52\linewidth}
\centering
    \centering
      \includegraphics[width=1\columnwidth]{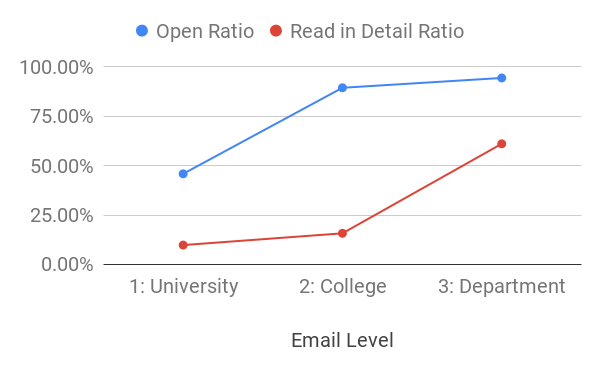}
      \caption{The Open/ Read in Detail Ratio of bulk emails of different levels collected in the interviews with recipients.}~\label{fig:open_level}
\end{minipage}\vspace*{-0.2in}
\end{table}

\subsubsection{The "organization" may well be fragmented.} Communicators usually were not involved in the transactional process of the messages. They were not responsible for the results of the organizational tasks in the organizational bulk emails. Thus the only performance metric used by communicators is open rate --- a ``proof of deliver'' to their clients. The organization's first priority --- getting the organization's tasks done --- was not valued in the ``open rates'' metric.

\subsubsection{Lacking of communications support technology.}
Communicators thought that personalization of email designs would be helpful but they did not have the technology to personalize subject lines and content. As both mailing lists and querying were limited to certain scenarios, precisely targeting was difficult for the communicators. To ensure the core population receives these messages, communicators usually sent bulk emails to a large community, much wider than the core targeting groups of the messages in these emails. 

\subsection{Conclusion}
The results of our study show that it is necessary to study organizational communication from multi-stakeholders' perspectives, as there are significant mismatches between the perspective of communicators who send out messages, recipients who receive them, and the management that represents the organization's interests. Therefore we propose the possible questions and research directions in studying remote-work organizational communication in the next section.

\section{A research agenda to study organizational communications in remote workplace}
\subsection{Key Questions}
A recent survey carried by Brynjolfsson et al. \cite{brynjolfsson2020covid} found that $34.1\%$ participants reported they were now working from home during April 2020. The transition from on-site work to remote work resulted in changes of communication channels. In this section we identify the key questions in this transition.

\subsubsection{What are the impacts of losing face-to-face communication channels?}
Organizational communication channels include face-to-face communication (informal chats, meetings, lectures) and virtual communication (instant messaging, emails, video conferencing, calls \cite{sullivan1995preferences}). Due to the change to the remote-work policy, face-to-face communication channels are greatly limited. The remained channels are carrying more tasks than before. For the messages which are used to be carried by face-to-face channels, which virtual channels are most appropriate to them? For the messages which are used to be sent by multiple channels and now losing some channels, will there be a large decrease in their effectiveness? How to design and distribute them such that we could still have the same effectiveness as before? 

\subsubsection{What communication channels are more effective in delivering which types of messages in remote-work environment?}
The messages we convey through organizational communication channels are also more important than ever. As we are switching to remote-work policy under a pandemic, we have lots of policy changes, new administrative operations, health information, which are both rapid and important. What are the recipients' retention of these messages via different communication channels is a problem unknown.

Some researchers have studied the performance of different communication channels \cite{fussell2000coordination} \cite{sullivan1995preferences} \cite{he2020remote,10.1145/3580507.3597677}. However, we don't know what mechanisms are or are not effective at reaching employees with central information in this work-from-home environment. Will employees read more or less bulk emails? Will they value the message from their local management? Will they attend to online meetings and join the discussion?  As society switches to the remote-work policy with an unprecedented large scale, what would the performance of communication channels be for the large-size organizations (for example, the University of Minnesota, with more than 50, 000 students and 25,000 employees) in the remote work situation? 

\subsubsection{What are the performance measures beyond information retention?}
The organizational communication in the remote-work is not only about information distribution and retention, but also about the creation of senses of connectedness, confidence, reassurance for employees. The creation of these senses are important because we are in a pandemic which asks for our collectivism to support shared-sacrifice. Organizations need to get their employees support on policies such like wear a mask for others, accept a furlough or temporary pay cut, and also get people to be happy with best-effort. As the virtual channels are more likely to cause misunderstanding compared to the face-to-face channels \cite{kraut2002understanding}, how to design the messages to appropriately create these senses and what channels should be used to convey these senses is an important question to be studied \cite{zhao2016group}. Also, as we knew from our prior work that employees already felt email overload in the onsite-work situation, will employees tolerate more emails in change of the sense of connectedness? How could we find a balance between making employees feel email overload and conveying the sense of connectivity?

\subsubsection{How can we align the different interests and perspectives of different stakeholders in the organizational communication systems? }
From our work in the organizational bulk email systems, we found that the organization might be fragmented. Recipients, managers, communicators, senders all have different opinions on what emails are important. Senders only want their words get out and want the communicators distribute them at the cost of communication channels \cite{kong2022multi, kong2023getting}.  Communicators usually are not involved in the transactional process of the messages and only want to get ``proof of deliver''. Recipients only care about the messages they are interested in and do not read administration messages from university leaderships while managers want their employees (the recipients) have some senses on what is going on in the university.

This fragmentation of organization might become a more serious problem in the remote-work situation \cite{kong2021virtual}. As in the remote-work situation, all organizational tasks which could be carried via face-to-face channels are now being communicated through virtual channels, whether the messages are really helping proceeding the tasks instead of being ``proof of deliver'' will decide the success of the organization goals. What are the role and value of communicator as an intermediary, and how to combine rhetorical analysis, management science in the design of remote-work organizational communication system needs to be studied.

\subsubsection{How can we design better technologies to support organizational communication in remote work environment?}
Communicators currently only have very limited technology to support them design and distribute messages. Recipients also lack technology to help them filter emails. As we have so many important messages to be sent out now (and to be received by recipients), the workload for communicators and recipients is becoming heavier than ever. Thus the technology support for message personizaton, targeting, distributing, feedback management, filtering is in great need.

\subsection{Potential Study Directions}
We proposed potential research around remote-work organizational communication systems here to fill in the gaps above:
\begin{itemize}
    \item Organization's goals and current practice on bulk emails in the remote work situation.
    
    \item The correlation between the content, amount of bulk emails with employees' sense of connectivity and email overload.
    
    \item Employee's time spent on bulk emails and consuming habits in the remote work situation.
    
    \item Employee's retention of content of message via different virtual communication channels.
    
    \item Employee's perspectives on bulk email values in the remote work situation.
    
    \item Specific designs to collect feedback and end-to-end performance metric on bulk emails.
    
    \item Natural-language-processing and machine-learning models \cite{kong2021nimblelearn,sun2023less} to predict email values to both the recipients and the organization; support communicators personalize and distribute messages according to these values; help recipients filter messages through hints of these values \cite{he2023hiercat, shen2022classifying, yi2020natural}.
    
    \item Apply personal information management strategies (PIM) \cite{lush2014fundamental} in the acquisition, organization, maintenance, and
retrieval of bulk email information.
    
\end{itemize}

\section{Conclusion}
The rapid change to remote-work policy brought many challenges to the organizational communication system. In this paper, we reviewed the prior work around evaluating, designing, and prototyping of organizational communications. We reviewed our recent findings and some research techniques we found useful in studying organizational communication. We identified the importance of using multi-stakeholders' perspectives and inbox-data. Last we identified the key questions of studying remote work organizational communications. We proposed a research agenda with potential study directions.

\section{Acknowledgments}

\bibliographystyle{ACM-Reference-Format}
\bibliography{sample-base}

\section{Appendices}
\appendix
\section{Structure of the Bulk Email System of the University of Minnesota}
\begin{figure}[!htbp]
\centering
  \includegraphics[width=1\columnwidth]{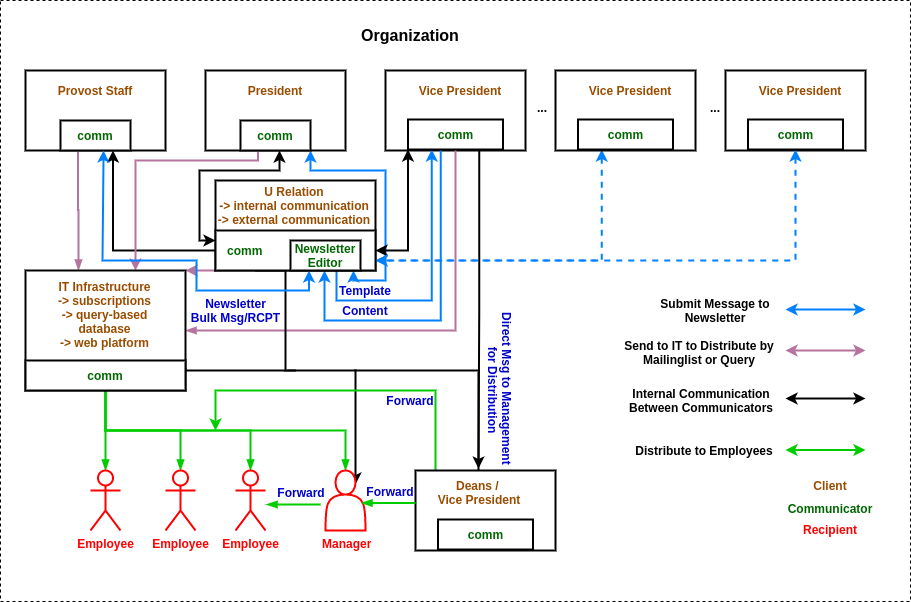}
  \caption{Structure of the Bulk Email System of the University of Minnesota; ``comm'' stands for communicators. }~\label{fig:comm}\vspace*{-0.28in}
\end{figure}
\end{document}